\documentclass[3p,twocolumn]{elsarticle}
\usepackage[colorlinks]{hyperref}
\usepackage{amsmath,amssymb,amsfonts,amsthm}
\usepackage{tikz}
\usepackage{pgfplots}
\newtheorem{definition}{Definition}
\newtheorem{lemma}{Lemma}
\newtheorem{example}{Example}
\newcommand{\refequ}[1]{Eq.~(\ref{#1})}
\journal{Information Processing Letters}
\begin{document}
	\begin{frontmatter}
		\title{Lower Bound Proof for the Size of BDDs representing a Shifted Addition}
		\author[a1]{Jan Kleinekath\"ofer}
		\ead{ja_kl@uni-bremen.de}
		\author[a2]{Alireza Mahzoon}
		\ead{mahzoon@informatik.uni-bremen.de}
		\author[a3]{Rolf Drechsler}
		\ead{drechsler@uni-bremen.de}
		\begin{abstract}
			Decision Diagrams(DDs) are one of the most popular representations for boolean functions. 
			They are widely used in the design and verification of circuits.
			Different types of DDs have been proven to represent important functions in polynomial space and some types (like Binary Decision Diagrams(BDDs)) also allow operations on diagrams in polynomial time.
			However, there is no type which was proven capable of representing arbitrary boolean functions in polynomial space with regard to the input size. 
			In particular for BDDs it is long known that integer multiplication is one of the functions, where the output BDDs have exponential size. 
			In this paper, we show that this also holds for an integer addition where one of the operands is shifted to the right by an arbitrary value. 
			We call this function the Shifted Addition. Our interest in this function is motivated through its occurrence during the floating point addition.
			
		\end{abstract}
	\end{frontmatter}
\section{Introduction}
As the demand for more complex circuits grows and advanced production techniques allow the production of such circuits, the tools for the design, verification and testing have to scale accordingly. 
Many of the corresponding tools are dependent one the efficient handling of boolean functions. 
Decision diagrams are among the most popular representations in those fields.
In particular Reduced Ordered Binary Decision Diagrams(ROBDD), usually just called BDDs, are often used.
It is long known that BDDs fail to represent the multiplication of two integers in polynomial space~\cite{B:1991a}. 
As a consequence of this result, efforts were made to overcome this limitation by adjusting the diagrams type.
Eventually, Multiplicative Binary Momentum Diagrams(*BMDs) were introduced and it was proven that they represent multiplication in polynomial space~\cite{B:1995a}. 
The trade-off for this capability is that the synthesis of *BMDs can take exponential time as stated in \cite{BDE:97}.
Accordingly, from the complexity perspective using BDDs is still advisable if the size of the BDDs is polynomially bound. 
Some results regarding the complexity of BDDs representing important are summarized in \cite{W:2004}.
The prove for the multiplication also implies that the square operation is not feasible. 
On the other hand for addition and some more function it is known how to represent them in polynomial space with BDDs.
In \cite{HY:1997} it is proven that BDDs cannot represent integer division in polynomial space.
For other functions, like the shifted addition, the complexity is still unclear.

In this publication we want to add a function with high importance for arithmetic circuits to the list of functions which are not representable in polynomial space with BDDs. 
The function of interest is an addition where one of the operands is shifted to the right by an arbitrary value.
Primarily, we are interested in this function because it occurs in floating point addition, when the significands are added after the alignment shift.
Many approaches for the verification of floating point adders found that the BDDs are getting extremely big for the result of the Shifted Addition and therefore use some mitigation strategy like i.e. case splitting ~\cite{CB:1998},~\cite{JWP:2005},~\cite{SJO:2005}.
To perform this lower bound proof we are using the concept of fooling sets.
The concept originates from VLSI design \cite{HJM:2018} and was applied for lower bound complexity of BDDs in~\cite{B:1991a}. 
\section{Preliminaries}
\subsection{Shifted Addition}
\refequ{sadd} represents the function of the shifted addition.
The two operands $A$ and $B$ are added and $B$ is shifted to the right by $D$.
If we want to access the $m$-th bit of an $n$-bit shifted addition we denote this as $sAdd^n_m$.
\begin{align}
	sAdd=A+(B\gg D) \label{sadd}
\end{align}

This function is a key part in the process of adding two floating point numbers. Floating point numbers are triples $(S,M,E)$ with $S$ being called the sign bit, $M$ the signifcand or mantissa and $E$ the exponent. 
The value of a floating point number is calculated as $-1^S\times M\times \beta^{E-B}$ where $\beta$ is the basis and $B$ a bias to allow negative exponents. 
The addition of two floating point numbers is performed by the combination on multiple integer arithmetic operations. 
First the significands of the two numbers have to be aligned and added and afterwards the result has to be normalized and rounded. 
\begin{figure}
	\includegraphics[width=0.5\textwidth]{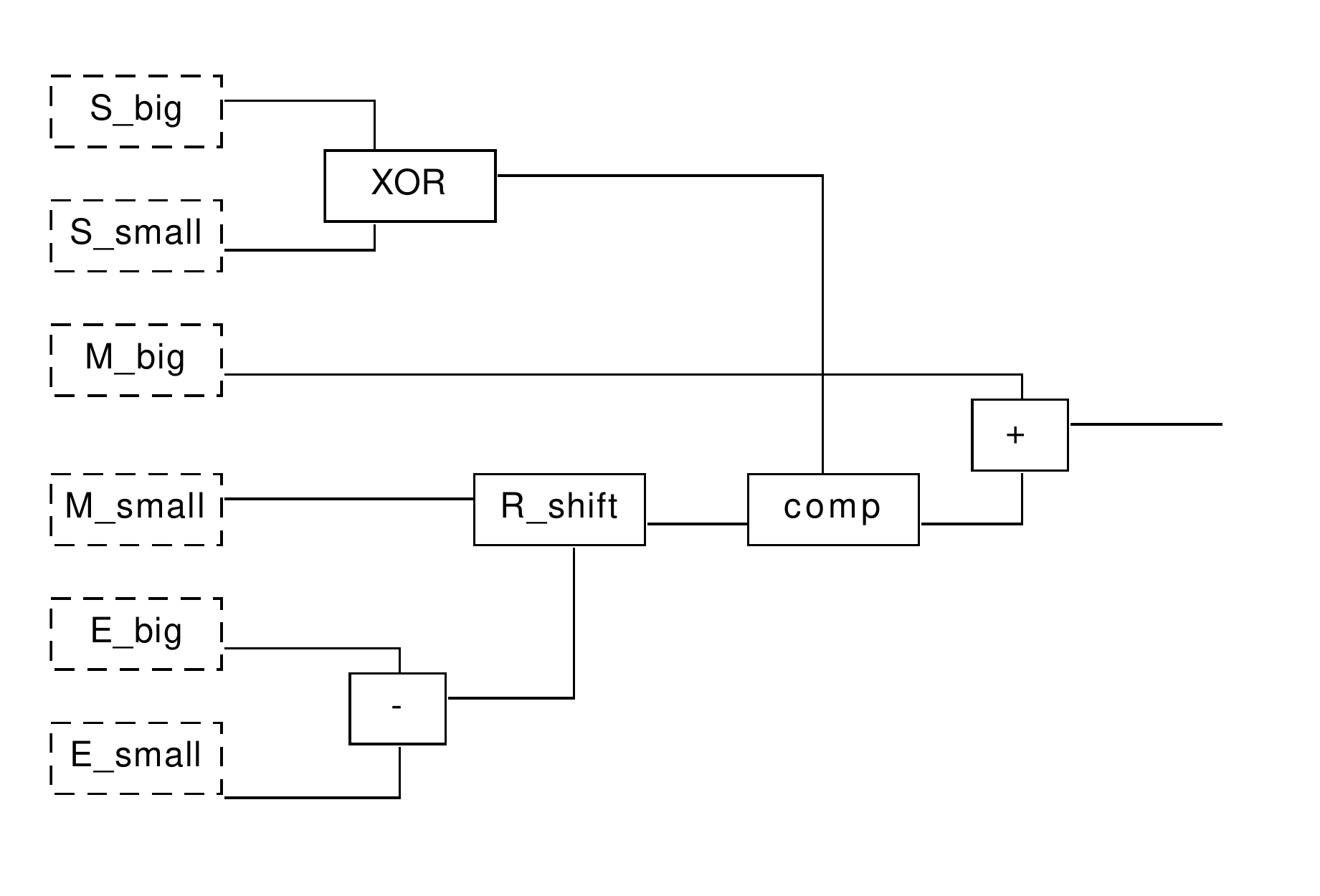}
	\caption{Addition of the significands of two floating point numbers\label{fig:fp_adder}}
\end{figure}
The first part can be seen in Figure \ref{fig:fp_adder}. 
Here it is assumed that it is already known which number is bigger. 
$E\_big$ therefore denotes the exponent of the bigger number. 
The difference of the two exponents is calculated by subtracting. The resulting difference will be the shift amount of the shifted addition. 
Next the smaller significand is shifted to the right by the exponent difference, which constitutes the shifted operand of the shifted addition. 
Before the actual addition takes places the shifted operand is possibly inverted, if the sign bits differ. 
When this happens the resulting function is a shifted subtraction instead. Finally the shifted operand and the significand of the bigger number are added. 
This generates the shifted addition. The rest of the addition with normalization and rounding is not important in this context.

\subsection{Binary Decision Diagrams}
BDDs are acyclic directed graphs which are used as a representation of boolean functions. In this paper we consider them in their reduced and ordered form like introduced in \cite{B:1986}.
They are suited for many applications because of three main properties: 
1)~there is only one BDD representation for a given function, 2)~ as shown in \cite{BRB:1990}, logical operations can be performed efficiently on BDDs, and 3)~there are many important functions (e.g.~integer addition), which can be represented by BDDs of polynomial size.

\subsection{Fooling sets}
Proofing lower bounds for BDD sizes of a given function is a difficult task, because it has to be proven that there is no variable ordering resulting in a smaller size. 
Consequently, it has to be argued over all of the exponentially many variable orderings.
The concept of fooling sets was introduced in~\cite{B:1991a} as a lower bound proof technique for BDD sizes to overcome this problem.
Instead of arguing over all possible variable orderings it can be argued about balanced partitions and it has to be shown that there is a set for every balanced partition that satisfies the properties of a fooling set.
To define fooling sets balanced partitions are used.
\begin{definition}
	A \emph{balanced partition} of $X$ into the sets $L$ and $R$ is defined by a subset $Y\subseteq X$ and $0\leq\omega\leq 1$ such that $\lfloor |Y| \times \omega \rfloor \leq |L| \leq \lceil |Y| \times \omega \rceil $ holds.
\end{definition}
\begin{example}
	Given the set $S=\{a_1,a_2,b_1,b_2\}$:
	\begin{itemize}
		\item $(\{a_1,b_2\},\{a_2,b_1\})$ is a balanced partition of $S$ with $\omega=0.5$.
		\item $(\{b_1\},\{a_1,a_2,b_2\})$ ---''--- $\omega=0.25$.
		\item $(\{a_1,a_2,b_1\},\{b_2\})$ ---''--- $\omega=0.75$.
	\end{itemize}
\end{example}

The balanced value $\omega$ is usually set to $0.5$.
We now define a fooling set based on the balanced partition definition.

\begin{definition}
	A \emph{fooling set} for a Boolean function $f$ and a balanced partition $(L,R)$ is a set $\mathcal{A}(L,R)$ which contains pairs $(l,r)$ of assignments. For two different pairs $(l,r),(l',r')$, it has to hold $f(l\cdot r)\not=f(l'\cdot r)$, where $l\cdot r$ denotes the complete assignments resulting from assignments $l$ and $r$.  
\end{definition}
\begin{example}
	\label{foolingsetexamp}
	Consider the function $f=a_2\oplus b_2 \oplus (a_1b_1)$ and the balanced partition $(\{a_1,b_1\},\{a_2,b_2\})$.
	We define $(l_1,r_1)=(\{0,0\},\{0,1\})$ and $(l_2,r_2)=(\{1,1\},\{0,0\})$ to be two assignments for this partition. 
	The set $\{(l_1,r_1),(l_2,r_2)\}$ is a fooling set.
	To show this the fooling set condition has to be checked:
	\begin{align}
		f((l_1,r_1)) &= 0 \oplus 1 \oplus (0\cdot0) &= 1 \label{l1,r1} \\
		f((l_2,r_1)) &= 0 \oplus 1 \oplus (1\cdot1) &= 0 \label{l2,r1}\\
		f((l_2,r_2)) &= 0 \oplus 0 \oplus (1\cdot1) &= 1 \label{l2,r2} \\
		f((l_1,r_2)) &= 0 \oplus 0 \oplus (0\cdot0) &= 0 \label{l1,r2}
	\end{align}
	As it can be seen in seen in Eq.\ref{l1,r1} and Eq.\ref{l2,r2} for both pairs the function evaluates to 1.
	When the left sites are swapped the result changes to 0 for both pairs as it can be seen in Eq. \ref{l2,r1} and Eq. \ref{l1,r2}.
	Accordingly the property of fooling sets holds. 
	This actually is the biggest possible fooling set for the given function and balanced partition, because the two left sides already cover $a_1b_1=0$ and $a_1b_1=1$. 
\end{example}

Although the definitions of balanced partitions and fooling sets are independent from the concept of BDDs, it can be used to argue about the complexity of BDDs for a given Boolean function. 
The balanced partition $(L,R)$ describes variable orderings where all variables from $L$ are above all variables from $R$. 
It can easily be seen that there is a balanced partition for every variable ordering.

\begin{example}
	Two of the variable orderings represented by the balanced partition from Example \ref{foolingsetexamp} are $(a_1,b_1,a_2,b_2)$, $(b_1,a_1,a_2,b_2)$.
\end{example}
The following lemma allows us to prove the exponential BDD size for all possible variable orderings.

\begin{lemma}
	If there is a fooling set with size $c^n$ and $n>1$ for every balanced partition $(L,R)$ for a function $f$; then, every BDD representing the function has a size of at least $O(c^n)$ .
\end{lemma}

While the full proof can be read in \cite{B:1991a}, let us give an intuition on why the lemma is correct.
Given an arbitrary variable ordering, there is the corresponding balanced partition $(L,R)$.
Directly below the last variable from $L$, the width of the BDD has to be at least as wide as the size of the fooling set. 
We call the number of nodes a BDD has for a variable the width for this variable and the overall width of the BDD than is the maximum width of all variables. 
Otherwise, there would be $l$ and $l'$ from the fooling set leading to the same node at this level of the BDD, which would result in a violation of the fooling set. It is due to the fact that no $r$ will be able to produce a different result for these $l$ and $l'$. 
Consequently, the complete BDD has to be bigger than the fooling set.

\section{Complexity of the Shifted Addition} 
Before beginning with the formal proof, we first highlight the key differences between the proof for the multiplication (presented in~\cite{B:1991a}) and our proof for the shifted addition:
\begin{itemize}
	\item Specification of the key variables: For the multiplication, the set of key variables is the first operand. We will instead define both operands as our key variables. 
	\item Integer alignment: While for the multiplication the positioning of two ones in the second operand is used to align the integers for the addition, for our shifted addition, we will use the shift value for the alignment. Consequently, the shift is only performed in one direction.
\end{itemize} 

We define the size of $A$ and $B$ as $n$, and $D$ has $m$ bits (see \refequ{sadd}). 
We assume that $m>log_2(n)$ always holds.
We define our set of key variables $Y=A\cup B$ and the weight of the balanced partition as $\omega=0.5$. 
According to \cite{B:1991a}, proving the following lemma automatically confirms that there cannot be a linear size BDD.
\begin{lemma}
	For every balanced partition $(L,R)$, the Boolean function $sAdd^n_{n-1}$ has a fooling set $\mathcal{A}$ such that $|\mathcal{A}|\geq 2^{n/4}$. 
\end{lemma}

\begin{proof}
	Let us first divide the set $Y$ into four sets as follows:
	\begin{align}
		A_L &= A \cap L \\
		A_R &= A \cap R \\
		B_L &= B \cap L \\
		B_R &= B \cap R
	\end{align}
	Since $A$ and $B$  as well as $L$ and $R$ have the same size, the sizes of the four subsets are strongly dependent on each other. If $A_L$ has size $k$, then the size of the other sets evaluates to $A_R=B_L=n-k$ and $B_R=k$.
	
	We use the value $p$ to denote the shift value, i.e.~$B$ is shifted to right by $p$ bits, where $0\leq p \leq n$.
	Moreover, the pairs of input bits from $A$ and $B$, which are going to be added, are denoted by $Args_p$. 
	
	\begin{align}
		Args_p = \{(a_i,b_{i+p}) \mid 0\leq i \leq n-p\}
	\end{align} 
	\begin{example}\label{exampargs}
		For $n>2$ $Args_2$ contains $(a_0,b_2)$.
	\end{example}
	
	To only get the pairs which are split between $L$ and $R$, the function $Split_p$ is defined.
	
	\begin{align}
		Split_p = Args_p \cap ((A_L\times B_R) \cup (A_R\times B_L ))
	\end{align}  
	\begin{example}
		Regarding Example \ref{exampargs}, $(a_0,b_2)$ is in $Split_2$ if $a_0\in L$ and $b_2\in R$ or if $a_0\in R$ and $b_2 \in L$ .
	\end{example} 
	It is now crucial to evaluate how many pairs are in $Split_p$ over all possible $p$. 
	Given $a_i\in A_L$ and $b_j\in B_R$, the pair $(a_i,b_j)$ contributes to a $Split_p$, if $j\geq i$. 
	The split, this pair is in, is $Split_{(j-i)}$.  
	The same argument is correct for pairs from $A_R\times B_L$.
	Given the set $(A_L\times B_R) \cup (A_R\times B_L )$, for more than half of the contained pairs $(a_i,b_j)$, $j\geq i$ holds. 
	\begin{align}
		\sum_{p=0}^n  |Split_p| &\geq \frac{|A_L\times B_R|+ |A_R\times B_L|}{2}\nonumber\\
		&= \frac{k^2 + (n-k)^2}{2}\\
		&\geq\frac{\frac{n}{2}^2+(n-\frac{n}{2})^2}{2}\nonumber\\
		&= \frac{n^2}{4}
	\end{align}
	This equation reaches its minimum for $k=n/2$. The sum of all shift values $p$ then evaluates to having at least size $\frac{n^2}{4}$. This also implies that there is at least one $p$ for which $Split_p\geq \frac{n}{4}$.
	From now on let us use the $p$ for which the size of the split is at least $n/4$. For this $p$, we will embed two integers $U=\{u_{m-1},\dots,u_0\}$ and $V=\{v_{m-1},\dots,v_0\}$ of the size $m=n-p$ in the input. $U$ is embedded in $A$ and $V$ in $B$.
	\begin{align}
		u_i = a_i \mid 0\leq i < m\\
		v_i = b_{i+p} \mid 0\leq i <m
	\end{align} 
	It is important to see that all pairs $(u_i,v_i)$ are contained in $Args_p$ and at least $n/4$ of them will be in $Split_p$.
	With all the prior considerations and definitions we can start to create the fooling set $\mathcal{A}(L,R)$. 
	The pairs $(l,r)\in\mathcal{A}$ only differ for pairs from $U$ and $V$ for which $(u_i,v_i)\in Split_p$.
	We start by defining how the rest of the inputs will look like:
	\begin{enumerate}
		\item All values in $A$ before $U$  are set to 1 to propagate the value: $x(a_i)=1 \mid \forall a_i\in A\setminus U$.
		\item All values in $B$ after $V$ are set to 0, because they have no influence on the calculation: $x(b_i)=0 \mid \forall b_i\in B\setminus V$
		\item The shift amount is set to $p$: $D=p$.
		\item For values from $U$ and $V$ which are not split between $L$ and $R$, fixed values are set. $x(u_i)=1,x(v_i)=0\mid (u_i,v_i)\in Args_p\setminus Split_p$
	\end{enumerate}
	\begin{align}
		\mathcal{A} = \{x \in \mathcal{X} \mid x(u_i)=\lnot x(v_i) \quad\forall (u_i,v_i)\in Split_p\}
	\end{align}
	The size of this fooling set is therefore dependent on the size of $Split_p$. With the already determined size of $Split_p$ it can be shown that the size of $\mathcal{A}$ is exponential.
	\begin{align}
		|\mathcal{A}| &= 2^{|Split_p|}\\
		&\geq 2^{n/4} \label{foolsize}
	\end{align}
	\begin{example}
		With the inputs $A=a_2,a_1$ and $B=b_2,b_1$ of size $n=2$ consider the balanced partition $(\{a_2,b_2\},\{a_1,b_1\})$. 
		$Split_1$ contains exactly $(a_1,b_2)$ and thus has size $1$. 
		$m$ evaluates to $m=2-1=1$. 
		According to the defined rules, for all pairs of the fooling set $a_2$ has to be $1$ and $b_1$ has to be $0$. 
		Moreover, it has to hold $a_1\not=b_2$.
		Therefore, the fooling set $\mathcal{A}$ will contain 2 pairs as presented in Eq. \ref{exampfoolset}.
		\begin{align}
			\mathcal{A} &= \{(\{1,0\},\{1,0\}),(\{1,1\},\{0,0\})\} \label{exampfoolset}
		\end{align}
	    The size of $\mathcal{A}$ also aligns with Eq \ref{foolsize}. $|\mathcal{A}| = 2 = 2^1$
	\end{example}
	
	To complete our proof, we now have to show that the pairs from $\mathcal{A}$ fulfill the requirements of a fooling set. 
	\begin{align}
		sADD_{n-1}^n=  \begin{cases}
			x(a_{n-1}) \oplus x(b_{n-1}) \\ \quad\quad x(u_i)\not= x(v_i) \text{ for all } 0<i<m-1 \\
			x(a_{n-1}) \oplus x(b_{n-1}) \oplus x(a_k) \\ \quad\quad k=max(i\mid  0 < i <m-1 
			\\\quad\quad\text{and } x(u_i) \not= x(v_i))  
		\end{cases}
	\end{align}  
	By definition, for all $x\in\mathcal{X}$ there is no $0\leq i<m-1 $ for which $x(u_i)=x(v_i)$ and the first case applies. By definition of the $\mathcal{X}$ either $x(a_{n-1})$ or $x(b_{n-1})$ is 1 and therefore $sADD^n_{n-1}(x)$ evaluates to $1$ for every $x\in \mathcal{X}$.
	
	In order to prove that $\mathcal{A}$ is a fooling set, we have to prove that we can find an $r$ for every two $l,l'$ such that $f(l,r)\not=f(l',r)$. 
	The idea is that there is a first point where $l_i\not=l'_i$ and this will produce a carry bit, which will be propagated to switch the output of the BDD to 0 for either $f(l,r')$ or $f(l',r)$.  
	The first important step into that direction is considering the possible alignments  of $u_i$ and $v_i$  into the partitions. 
	These alignments are presented in TABLE \ref{uvlocations}. 
	When both variables are in the same partition they have static values as described in the definition of $\mathcal{X}$. 
	If they are split between the partitions, they are defined to have alternating values. 
	When it is looked closer at the relation between $l\cdot r'$ and $l' \cdot r$ it can be observed that $l \cdot r' (u_i) = \lnot l' \cdot r(v_i) $ as well as $l \cdot r' (v_i) = \lnot l' \cdot r (u_i)$.
	\begin{table}
		\begin{center}
			\caption{Locations of $u_i$ and $v_i$\label{uvlocations}}
			\begin{tabular}{|c|c|c|c|c|c|c|c|c|c|}
				\hline
				\multicolumn{2}{|c|}{Location} & \multicolumn{2}{|c|}{$l\cdot r$}  &  \multicolumn{2}{|c|}{$l'\cdot r'$} &  \multicolumn{2}{|c|}{$l\cdot r'$}  &  \multicolumn{2}{|c|}{$l'\cdot r$}  \\
				\hline
				$u_i$& $v_i$ & $u_i$ &  $v_i$ & $u_i$ &  $v_i$ & $u_i$ &  $v_i$ & $u_i$  &  $v_i$ \\
				\hline
				\hline
				L & L & 1 & 0 & 1 & 0 & 1 & 0 & 1 & 0 \\
				\hline
				R & R & 1 & 0 & 1 & 0 & 1 & 0 & 1 & 0 \\
				\hline
				L & R & $a$ & $\lnot a$ & $b$ & $\lnot b$ & $b$  & $\lnot a$ & $\lnot b$ & $a$ \\
				\hline
				R & L & $a$ & $\lnot a$  & $b$ & $\lnot b$ &  $\lnot b$ & $a$  &  $b$  & $\lnot a$  \\
				\hline
			\end{tabular}
		\end{center}
		
	\end{table}
	Two possible cases have to be considered:
	\begin{enumerate}
		\item $a_{n-1}=b_{n-1}$ and $u_i\not= v_i $ for all $0\leq i < m-1$ if $p=0$ or $0\leq i <m$ otherwise: 
		\begin{align}
			sAdd^n_{n-1}(x)&=x(a_{n-1})  \oplus x(b_{n-1}) \\
			&= 0
		\end{align}
		\item Otherwise there has to be a $k$ with $k=max(i\mid  0 < i <m-1 \text{ and } x(u_i) \lnot x(v_i))$:
		\begin{align}
			&sAdd^n_{n-1}(l\cdot r') \\
			&= l\cdot r'(a_{n_1}) \oplus l\cdot r'(b_{n-1}) \oplus l \cdot r'(u_k) \\
			&= \lnot l'\cdot r (b_{n-1}) \oplus \lnot l'\cdot r(a_{n-1}) \oplus \lnot l' \cdot r(u_k)\\
			&= \lnot (l'\cdot r (a_{n-1}) \oplus  l'\cdot r(b_{n-1}) \oplus  l' \cdot r(u_k))		\\
			&= \lnot sAdd^n_{n-1}(l'\cdot r)	
		\end{align}
		This means that either for $l\cdot r'$ or $l'\cdot r$ the result will be 0, which is the opposite of $l\cdot r$ and $l' \cdot r'$. 
	\end{enumerate}
	This finally proves that the set is a fooling set and due to its exponential size, there cannot be a BDD with polynomial size.	
\end{proof}
\section{Experiments}
\begin{figure}
	\begin{tikzpicture}
		\pgfplotsset{width=.35\textwidth,height=.3\textwidth,compat=newest}
		\begin{axis}[
			scale only axis=true,
			y label style={at={(-0.18,0.5)}},
			xlabel={ \small Significand size },
			xmin=2, xmax=16,
			ymin=0, ymax=100000,
			ymode=log,
			xmode=log,
			xtick={2,4,8,16},
			xticklabel={%
				\pgfmathparse{2.718281828^\tick}\pgfmathprintnumber\pgfmathresult
			},
			legend pos=north east,
			]
			
			\addplot[
			mark=square,
			color =blue 
			]
			coordinates {
				(2,5)(4,16)(8,136)(16,5851)
			};
			
			\addplot[
			mark=square,
			color =red 
			]
			coordinates {
				(2,1)(4,2)(8,8)(16,128)
			};
			\addplot[
			mark=square,
			color = green  
			]
			coordinates {
				(2,22)(4,96)(8,937)(16,46761)
			};
		\end{axis}
	\end{tikzpicture}

\caption{red: predicted lower bound, blue: observed BDD width, green: observed BDD size\label{fig:experiment}}
\end{figure}
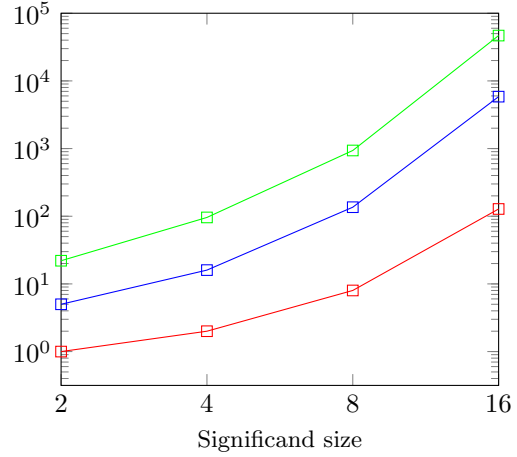
To evaluate the result experimentally, BDDs representing shifted addition were build for operand sizes of 2, 4, 8 and 16. 
Because the perfect variable ordering for the shifted addition is unclear an for every operand size 50 random variable orderings were tested and only the minimum sizes are reported here. 
The results are presented in Figure \ref{fig:experiment}. 
In addition to the size of the BDDs (in green) also the width (in red) is reported as the fooling set proof sets a lower bound for the width which only implies a lower bound for the size. 
Furthermore, the lower bounds proven in this paper are presented in blue.

First of all it can be seen, that both reported sizes stay above the predicted lower bound for all operand sizes. 
As it can be expected, the BDD size always is significantly bigger than the BDD width.
In addition for the operand size of 16 it can be seen that the width of the BDD with 5851 is far bigger than the predicted lower bound of 128. 
\section{Conclusion}
With the help of fooling sets we were able to show that BDDs cannot be used to represent Shifted Addition in polynomial space. 
This backs the observation that was made by different authors during the verification of floating point adders. 

Studying the class of functions that are representable by BDDs in polynomial space is interesting beyond the Shifted Addition as explosions in the size of BDDs are often observed and knowing whether this explosion is due to a bad variable ordering or due to the represented function is a valuable insight.

Further research regarding the lower bound for shifted addition with regard to BDDs could involve the application of different proof techniques to improve the bound.
Historically, this was achieved for binary multiplication after a first proof with fooling sets was introduced \cite{BW:2001}.

\bibliography{lit_header_long,ipl_paper,lit_focus,lit_my,lit_others}
\end{document}